\documentclass[12pt]{amsart}   

\usepackage{graphicx}
\usepackage{tikz}
\usepackage{amssymb} 
\usetikzlibrary{calc}
\usepackage{url}
\usepackage{caption}
\usepackage{tcolorbox}
\usepackage{listings}

\theoremstyle{definition}

\theoremstyle{remark}

\title{Top-down Automated Theorem Proving\\(Notes for Sir Timothy)
}

\author{C. E. Larson and N. Van Cleemput}
\address{Department of Mathematics and Applied Mathematics\\Virginia Commonwealth University\\Richmond, VA 23284, USA }
\address{Department of Applied Mathematics, Computer Science and Statistics\\Ghent University\\Ghent, Belgium }
\thanks{(*) Research supported by the Simons Foundation Mathematics and Physical Sciences--Collaboration Grants for Mathematicians Award (426267)}

\date{}

\begin{document}

\begin{abstract}
We describe a ``top down'' approach for \textit{automated theorem proving} (ATP). 
Researchers might usefully  investigate the forms of the theorems  mathematicians use in practice, carefully examine how they differ and are proved in practice, and code all relevant domain concepts. These concepts encode a large portion of the knowledge in any domain. Furthermore, researchers should write programs that produce proofs of the kind that human mathematicians write (and publish); this means proofs that might sometimes have mistakes; and this means making inferences that are sometimes invalid. 

This approach is meant to contrast with the historically dominant ``bottom up'' approach: coding fundamental types (typically sets), axioms and rules for (valid) inference, and building up from this foundation to the theorems of mathematical practice and to their outstanding questions.
It is an important fact that the actual proofs that mathematicians publish in math journals do not look like the formalized proofs of Russell \& Whitehead's \textit{Principia Mathematica} (or modern computer systems like \textsc{Lean} that automate some of this formalization). We believe some ``lack of rigor'' (in mathematical practice) is human-like, and can and should be leveraged for ATP.
\end{abstract}

\maketitle










\section{Background}

 In 1948 Turing \cite{Turi48} suggested mathematics as a subject that intelligent machines might contribute to, and in 1958 Newell and Simon \cite{NeweSimo58} predicted, ``That within ten years a digital computer will discover and prove an important new mathematical theorem.'' More than 60 years later, computer proof of an important new theorem still seems distant. What might be wanted is a program that, given a conjecture in the domains research mathematicians work in (like number theory, matrix theory, graph theory, etc) as input,  sometimes produces a proof of the conjecture. Nothing like this exists. There have been scattered computer proofs of conjectures, most famously the 1997 EQP proof of the Robbin's conjecture \cite{Mccu97}, and lots of related research, but no track record yet to build on. Some \textit{automated theorem proving} (ATP) researchers have begun looking for new ideas and approaches\footnote{This note was inspired by a blog post of the Fields' medalist Timothy Gowers, who has written about his attempts to code ``human-style'' reasoning for ATP. \textit{To err is human.}}.
Here we propose a ``top-down'' approach: develop programs that produce proofs in the domains that mathematicians work in. This approach (``top-down ATP'') is meant to contrast with the more traditional  bottom-up approach: develop programs that produce proofs from axioms of logic or set theory (or foundational relatives, such as dependent type theory); after this is done, translate any conjecture in an existing mathematical domain to the language of set theory.

The bottom-up ATP approach has been dominant. Logicians and set theorists in the first half of the 20$^{th}$ century argued variously that mathematics was in fact logic (that is, that mathematical statements were logically true) and that all of mathematics could be derived from various set-theoretic axioms and exactly specified inference rules. The first automated-theorem proving programs in the 1950s in fact aimed to prove the theorems of Russell and Whitehead's systematization of mathematics \textit{Principia Mathematica} \cite{RussWhit97}; these programs included, for instance, Newell and Simon's \textsc{logic theorist} program \cite{NeweSimo56} and Wang's IBM-era programs \cite{Wang60}. Well-known advances in this area of research included simplified or efficient inference rules such as \textit{resolution theorem proving} \cite{Robi65}. It is worth noting that proof-assistants like \textsc{Lean}---which are not intended to prove theorems automatedly---which leverage computers in the task of formalization---are also bottom-up: computers can be used for type-checking, making routine steps and keeping track of dependencies and managing large databases, but they start with bottom-level foundational objects, and adhere to strict inference rules.


At the same time there has been continuing work on top-down ATP approaches that did not emphasize reduction to logic or set theory, or that were domain-specific. The first of these might have been Gelerntner's geometry program \cite{Gele59}. Some of this research in fact has been spectacularly successful. Wu's work on geometry is one example \cite{Wu12}. A less-recognized example is WZ theory and Zeilberger's work on hypergeometric series: huge classes of identities, including many famous and classical identities, can now be proved completely automatedly \cite{WilfZeil92}. Top-down ATP research might have had limited impact among bottom-up ATP researchers exactly because it was not obvious how this research could advance goals of researcher with less specific, more general, theorem-proving ambitions.

There is now a recent burst of energy and publicity for a variety of efforts including proof checking programs, interactive theorem-provers, and proof-assistants, which are relatives of bottom-up ATP \cite{Orne20}. There is a very long history of proof-checking programs; the \textsc{COQ} program has received generous coverage in the mathematical community at least in part because of the advocacy of Voevodsky \cite{Hart15}, checking proofs of many famous theorems. The \textsc{Lean} program has received substantial coverage, in part because of its adoption by working mathematicians (and the millions of lines of coded mathematical theorems its users have contributed). It began at Microsoft Research (see: \url{https://leanprover.github.io/about/}), and  Google Research  is also now participating \cite{RabeSzeg21}---substantial institutional support.

The terms top-down ATP and bottom-up ATP are meant to suggest an analogy with top-down programming versus bottom-up programming. ``Programming 
is traditionally taught using a bottom-up approach, where details of syntax and implementation
of data structures are the predominant concepts. The top-down approach proposed focuses instead on understanding the abstractions presented by the classical data structures without regard to their physical implementation'' \cite{Reek95}.
In top-down programming you begin by outlining top-level functions and add auxiliary functions, substructures, and basic tools as you discover you need them; while in bottom-up programming you first code the most fundamental modules first and build up from them. It may be that bottom-up approaches to ATP end up focusing disproportionately  on fundamentals and rigor, rather than on producing proofs. One advocate for top-down programming has written, ``We choose to focus on the [high-level] abstractions in the belief that they are the more important concepts. Our experience has been that implementation issues distract the students to the point that they do not really understand the abstractions, particularly with the classical data structures'' \cite{Reek95}.
We envision top-down ATP proofs to typically have holes, and filled-in and  fixed-up as needed; while bottom-up ATP starts with some fundamental objects, and inference rules, and builds up to some area of interest to mathematicians without gaps and always validly. 
It is reasonable to expect that top-down and bottom-up approaches to ATP can interact fruitfully---as they do in programming generally.

Furthermore we propose writing programs that produce proofs which are importantly similar to the proofs which appear in published journal articles: published proofs are generally valid, but \textit{have occasional mistakes}---and thus these are \textit{not} proofs (in general) which are derived from axioms via (valid) rules of inference. Published proofs of course \textit{do} have mistakes: Kempe's proof of the Four Color Theorem is a famous historical example  \cite{BiggLloyWils86}. The Fields Medalist Vladimir Voevodsky discussed several errors (including his own) in published papers \cite{Voev14}. Of course, it might seem to be a great advantage that we can produce errorless computer proofs---but perhaps commitment to this ideal has been a constraint on other automated theorem-proving goals?

\section{Examples}

We sketch the outline of a well-known theorem in graph theory in order to illustrate ideas. While it is the form of the argument that is of interest, we define enough terminology that a reader who has seen this theorem before might recall or reconstruct the proof. 

The \textit{degree} of a vertex in a graph is the number of vertices it is adjacent to. A \textit{Hamiltonian cycle} in a graph is a cycle that contains all the vertices of the graph. Let $n$ be the number of vertices in the graph. Dirac's Theorem says that if every vertex has degree at least $\frac{n}{2}$ then the graph has a Hamiltonian cycle. The main idea of the standard proof of this theorem is to  consider a longest path in a graph, argue that the subgraph induced by these vertices is itself Hamiltonian, and then to argue that this subgraph contains all the vertices of the parent graph. It is important that the claim has the form: $\forall x [P(x)\rightarrow Q(x)]$, where the quantification is over all \textit{graphs}, where $P(x)$ represents ``graph $x$ has the property that every vertex of $x$ has degree at least $\frac{n}{2}$'' and $Q(x)$ represents 		``graph $x$ has the property that $x$ is Hamiltonian''. 

In this case one approach to proving Dirac's Theorem would be to find an appropriate graph property $P'$ and predicate $P'(x)$ representing 	``graph $x$ has property $P'$'' and proving both $\forall x [P(x)\rightarrow P'(x)]$ and $\forall x [P'(x)\rightarrow Q(x)]$. The main question then is how to \textit{automate} finding or producing the needed property $P'$ (in this case we know it exists; while in the general case this would be an open question).  Let $R(x)$ represent ``the vertices in every longest path in graph $x$ induce a subgraph of $x$ which is  Hamiltonian'', and let $S(x)$ represent ``every longest path in  graph $x$ contains all of the vertices of the graph''. So then the proof of Dirac's Theorem has the form:
\[
\forall x [P(x)\rightarrow R(x)],
\]
\[ \forall x [(P(x) \& R(x)) \rightarrow S(x)],
\] 
\[
\forall x [(P(x) \& R(x) \& S(x)) \rightarrow Q(x)],
\]
 and we conclude: 
 \[
 \forall x [P(x)\rightarrow Q(x)].
 \]
Expanding the predicates, this says: (1) For every graph $x$, if $x$ has the property that every vertex of $x$ has degree at least $\frac{n}{2}$, then the vertices in every longest path in graph $x$ induce a subgraph of $x$ which is  Hamiltonian. (2) For every graph $x$, if $x$ has the property that every vertex of $x$ has degree at least $\frac{n}{2}$, and the vertices in every longest path in graph $x$ induce a subgraph of $x$ which is  Hamiltonian, then every longest path in  graph $x$ contains all of the vertices of the graph. (3) For every graph $x$, if $x$ has the property that every vertex of $x$ has degree at least $\frac{n}{2}$, the vertices in every longest path in graph $x$ induce a subgraph of $x$ which is  Hamiltonian, and every longest path in  graph $x$ contains all of the vertices of the graph, then graph $x$ has the property that $x$ is Hamiltonian. (4) Therefore, for every graph $x$, if $x$ has the property that every vertex of $x$ has degree at least $\frac{n}{2}$, then  $x$ has the property that $x$ is Hamiltonian.

It is worth noting that $P(x) \& R(x)$ and $P(x) \& R(x) \& S(x)$ represent \textit{new} predicates (of course constructed by conjunctions of existing predicates---and which can be given a simple new name should that be thought useful); this shows the sought-after predicate $P'(x)$ can be taken to be $P(x) \& R(x) \& S(x)$. So one approach to automating finding proofs of theorems is to focus specifically on the domain of the claim, focus specifically on how theorems of a specified form are actually proved, and produce candidate statements that utilize and leverage all the existing knowledge (conceptual knowledge and theorems) in that domain. 

This proof is a top-down proof as it relies on high-level abstractions (graphs)---not redefined in terms of low-level abstractions (sets)---and a library of graph concepts, together with high-level inferences which are not themselves directly justified in terms of low-level inference rules. There are three possible responses to such a produced ``proof''. A mathematician might accept the proof. Or she might request a justification for the intermediate claims. Or she might have a counterexample for one of these claims. It is worth noting that any of these three responses can advance mathematics. In the first case, we have a new theorem. In the second case, our mathematician will have to go back and get to work to provide the needed justification, perhaps inventing new concepts. And in the third case, counterexamples are themselves new knowledge  (and the numerous books  with titles like \textit{Counterexamples in Analysis} \cite{GelbOlms03} attest to their role in  fertilizing mathematics).

Below we will explain how ideas from  a program we developed to make conjectures in any mathematical domain  might be leveraged to produce proofs like this. The key features of our \textsc{Conjecturing} program are state-of-the-art expression generation, and an intuitive  heuristic for conjecture production. Produced conjectures are not directly inferred via syntactic rules, but rather satisfy semantic conditions. In particular, these conjectures must be true for the (possibly very small number of) examples known to the program. Wos, for instance, claimed that,
``An emphasis on semantics rather than syntax has far greater potential for producing a dramatic impact on the power of automated reasoning programs'' \cite{Wos88}. 
It is worth noting that the conjectures produced by this program have richer semantic content than the statements produced by many proof-assistant programs:  \textsc{Conjecturing} statements (conjecture \textit{objects}) come with \textit{methods} for instance for evaluating the truth of the conjecture claim for any given object of the appropriate type. 

This example was chosen to be sufficiently non-trivial, and enough to suggest an approach that can be generalized to other cases---but more difficult cases and the issues they raise will be discussed below. A key question, to be addressed, is \textit{where} the produced predicates come from (is there a library of all possible predicates, or are novel predicates somehow  generated for specific problems) and \textit{how} these predicates are actually produced (how does one get chosen over another)?

 Importantly, if one mathematician were to explain the proof to another, no explicit mention will be made of inference rules, and a common set of concepts will be relied on. The above  proof may or may not suffice. If it doesn't, further explanation will be required. This may involve additional concepts and properties. If a mathematician thinks that an inference isn't valid, she can supply a counterexample.

Consider now Gowers' example \footnote{\url{https://wtgowers.github.io/human-style-atp/2022/09/09/basicalgorithm.html} (accessed  July 2023)}: he discusses how a human-like automated theorem prover might prove the theorem that, in a metric space $X$, if $A$ and $B$ are closed sets then their union $A\cup B$ is closed. A ``metric space'' might be thought of in at least two ways: first, as a mathematical object in its own right, or second, as a generalization of specific metric spaces such as $\mathbb{R}^2$ (the Cartesian plane). While mathematicians have no problem reasoning about an abstract metric space, these are qualitatively different than specific examples of metric spaces. 

A graph can be represented in a variety of ways, graph properties can be coded, and whether a graph has a specific property can be checked. It is less obvious how to represent an abstract metric space---how for instance can one be coded, or their sets be coded, or the properties their sets may have be coded and evaluated? It may be an important fact that only very initial steps have been taken thus far in the widely-used mathematical computation environment \textsc{Sage} (in contrast there are well-developed facilities for graph, integers, matrices and more prosaic mathematical objects). Thus a useful first step in investigating Gowers' theorem-of-interest is to take up a specific case like $\mathbb{R}^2$. Future investigations of how to represent an abstract metric space to a computer might lead to a better understanding about how to prove facts about abstract metric spaces. It may be an important fact that properties about graphs apply to specific graphs---and not to an abstract ``graph'' object; and it might then be thought that metric space properties only apply to specific metric spaces (like $\mathbb{R}^2$) rather than an abstract metric space.

Consider now how  a human-like automated theorem prover might prove the theorem that, in the metric space $\mathbb{R}^2$, if $A$ and $B$ are closed sets then their union $A\cup B$ is closed.
In the case of  $\mathbb{R}^2$, it is much clearer how to represent relevant concepts to a computer:  $\mathbb{R}^2$ consists of points $(x,y)$ where $x$ and $y$ are real numbers. Sets of points in  $\mathbb{R}^2$ can be represented in various ways: either by listing specific points or by giving defining conditions; and algorithms can be written to determine if at least some of these sets are \textit{open} or \textit{closed}, etc. The objects of interest here are actually \textit{pairs} of sets in  $\mathbb{R}^2$. Properties here might include, for instance, the property $P$ that every set in the pair is closed, and the property $Q$ that the union of the sets is closed; corresponding predicates would be $P(x)$ representing, ``for a pair of sets x=(A,B), $A$ and $B$ are each closed'', and $Q(x)$, representing,   ``for a pair of sets x=(A,B), their union $A\cup B$ is closed''.

It may be that the program library includes the predicate $P'(x)$ representing ``for any pair of sets x=(A,B) and points $p_A\notin A$, $p_B\notin B$,  there are real numbers $\epsilon_A$ and $\epsilon_B$,  and balls $\mathcal{B}(p_A,\epsilon_A)$ disjoint from $A$ and $\mathcal{B}(p_B,\epsilon_B)$ disjoint from $B$''. And it may also include the predicate $P''(x)$ representing ``for any pair of sets x=(A,B)  and point $p \notin A\cup B$, there is a  real number $\epsilon$ and ball $\mathcal{B}(p,\epsilon)$ disjoint from $A\cup B$''. Then a produced proof might have the form: $\forall x [P(x)\rightarrow P'(x)]$,  $\forall x [(P(x)\& P'(x))\rightarrow P''(x)]$,  $\forall x [(P(x)\& P'(x)\& P''(x))\rightarrow Q(x)]$ and, therefore  $\forall x [P(x)\rightarrow Q(x)]$.

While the suggested top-down approach may seem unrealistic, in fact existing conjecture-making software can be leveraged to initiate the proposed research. Fajtlowicz for instance developed the \textsc{Graffiti} program in the 1980's  to produce graph-theory conjectures of the described form \cite{Fajt88a,Fajt88,Fajt87,Fajt90,Fajt95}. His heuristic ideas were domain-independent (that is, not specific to graph theory); more recently the \textsc{Conjecturing} program was developed to extend Fajtlowicz's \textit{Dalmatian} heuristic for the  production of conjectures of the described form in any mathematical domain \cite{LarsVanc16,LarsVanc17}.


 There are some obvious objections to the proposed top-down ATP paradigm. We address these immediately and then return to fleshing out possible top-down approaches.

\section{Issues and Objections}

A number of issues can be anticipated. Many of these simply capture that humans (even mathematicians) use words in different ways; in every instance it is important to understand what definition is being used (or what concept the word is meant to name). A logician's conception of a ``proof'' for instance differs substantially from a journal editor's conception. (While they may claim to actually have the same conception, what they \textit{count} as a ``proof'' differs in practice.) Lakatos' \textit{Proofs and Refutations} \cite{Laka76} is a key source for important discussions of historical examples occurring in mathematics; he discusses for instance the evolving concepts associated with ``polyhedron'' and ``convex set'' in the context of a discussion of mathematicians understanding of the ideas and proof of Euler's Polyhedral Formula (that the number of vertices of a polyhedron plus its number of faces equals two plus its number of edges.) These issues aren't often discussed among mathematicians---but are certainly common.

\begin{enumerate}
\item \textbf{It might be claimed that a proof consists of valid inferences from proceeding statements.} Words in natural languages and even mathematical languages are used in multiple ways. The word ``graph'' for instance is used both for \textit{graphs} of quadratic functions and \textit{graphs} in graph theory (while these terms arguably have \textit{some} commonality, it is easy to see that they are importantly different: it would make no sense, for instance, to ask how many edges the graph of a quadratic function has). Similarly, the word ``proof'' is used variously. In one context it means a sequence of statements inferred from axioms by specified rules; in contrast a ``proof'' as it appears in a math journal means a sequence of statements validly inferred from some assumptions. Here ``validly inferred'' \textit{cannot} mean according to any specified rules of inference as these are never specified (say in the ``author instructions'' for the journal), and as such ``proofs'' do not exist in any mathematics journal (and would be instantly rejected). By ``proof'' we mean something closer to what mathematicians actually submit to journals than the ``proofs'' of formal logic. It might be claimed that the ``proofs'' in mathematical journals can in fact be translated to formal ``proofs''; this is an open question---and beside the point. \\


\item \textbf{It might be claimed that produced proof statements are not justified unless they are valid inferences.} If ``justified'' is \textit{defined} to mean ``follows from axioms by specified inference rules'' then no published proofs are ``justified''. More interestingly, in an important sense even formal proofs in logic are not ``justified''---at some point, in any proof---steps must be made which cannot be justified (that is, there must always be an unjustified inferential leap).


The issue here was recognized at least by the 19$^{th}$ century and is succinctly described in Lewis Carroll's story of Achilles and the Tortoise \cite{Carr95}. Achilles knows the \textit{modus ponens} inference \textit{schema}: given the statement forms $\mathcal{A}\rightarrow \mathcal{B}$, and $\mathcal{A}$ then imply $\mathcal{B}$; but in the case of specific statements, ``if $P$ then $Q$'' and ``$P$'', how is the \textit{application} of the inference schema justified? The gist is that at some point we have to go from one statement to the next without any justificatory rule. (We can often give justifications, but at some point our justifications necessarily come to an end). 

Not every statement in a math paper can be justified. \textit{At some point} there can be no further justification. If the reader keeps pushing there is necessarily a limit.  So even in a state-of-the-art program that proves theorems using just formal inference rules, with a simple user-input of  ``All connected graphs have exactly one component'' and ``The Petersen graph is connected'', the program might output, ``The Petersen graph has exactly one component''. The user might wonder how the program can justify that inference. It might be said that the program has the rule for universal instantiation (UI), the input universal statement is a claim about ``all graphs'' and the Petersen graph is certainly a graph, so the program produced the inference about the Petersen graph. But how can the application of the UI schema be justified in \textit{this case}? Another rule would seem to be required: given the UI schema and appropriate statements in the domain of graphs, make an inference about any particular graph, etc. 

Of course, no mathematician wrestles with these endless justifications in practice: we all go along the same way here, agreeing on the same basic inferences. But that's an important fact. This is really what a human-oriented theorem proving program must \textit{model}. To be human-oriented must mean something like approximating what human mathematicians \text{do}. This is not meant to be an obscure philosophical point---but maybe a central design idea: a human-oriented theorem-proving program will make inferential leaps, not always justified, and sometimes invalid. What this fact highlights is that these programs should also then have some mechanism so that they can be corrected and not make the same mistake again. (That's what \textit{humans} do).

No theorem-proving program can expect to do better---the best they can do is produce sequences of statements where some (or most) human mathematicians think the inferences are justified, and in general, humans take as justification far less than long sequences of statements from axioms and logical rules. It may even be a central design feature to make inferential leaps with a mechanism for correcting mistakes and not repeating invalid ones in the future.\\

\item \textbf{It might be claimed that a theorem-proving program must produce ``proved'' statements that are inferred validly.} What is being proposed is to develop programs that produce ``proofs'' of the kind that human mathematicians produce. It is true that this is different---and necessarily different---than the bottom-up theorem-proving programs where the majority of research has been invested. The word ``proof'' can variously mean the proofs of formal logic---or the proofs published in math journals. These are sometimes different. The first are valid by definition, while the second are only ideally valid. Producing proofs of both sorts can be interesting, useful, and illuminating.\\

\item \textbf{It might be claimed that, since every mathematical statement can be translated into a statement about sets or other fundamental objects, that an automated theorem proving program should be built from some choice of fundamental object.}
Graph theory papers, for instance, generally prove statements about  \textit{graphs} directly as graphs (and not as sets of pairs satisfying some conditions); when researchers write graph theory algorithms, they write algorithms directly for graphs (and not algorithms for  sets that model graphs); and claims about graphs are written in the vocabulary of graph theory. While mathematical objects all can be modeled as sets, in practice mathematicians don't actually switch from talking about their objects of immediate interest to talking about the sets that can model these objects. There are a few reasons. These certainly include that statements in the home domain are simpler, and mathematicians have more developed intuitions about the objects of their home  domain. 

So while it is of course true that working only in set theory would be simplifying in the sense that only a single mathematical domain would have to be coded for the purposes of producing proofs, any theorem-prover that is to attain the goal of inputting a domain mathematical statement and occasionally getting a proof out will still need to translate that statement into set theoretic language and thus isn't necessarily any simpler. More importantly, the translated statements will almost certainly be longer and more complex---and likely harder for a theorem-proving program to prove. There are advantages to working with the objects of the various mathematical domains---and is the reason mathematicians themselves don't translate their conjectures and theorems to set theoretic claims.

The claim that mathematics is all one domain is of no practical import: graph theorists will study graphs, number theorists will study integers, and in so far as there are analogies, or common tools, the necessary bridges will be built. No mathematician for instance would ever attempt to prove a graph theoretic claim by translating it into a set theoretic claim and prove it directly from some axioms for set theory.

\item \textbf{It might be asked which axioms should be used in a theorem-proving program?} Questions like this are likely related to debates about which set theoretic axioms are true. They are no more (or no less) relevant to program design than they are to the practice of any mathematician. In fact, most mathematicians are agnostic about the truth of any specific collections of set theoretic axioms, and these don't generally come up in the practice of mathematical domains besides set theory. If you are developing a top-down mathematical theorem-proving program there may be design reasons to include theorems (facts) from that domain. These in a sense will then be the ``axioms'' for the program.

\end{enumerate}

\section{Human-style ATP and the \textsc{Conjecturing} Program}

The Fields Medalist Timothy Gowers recently returned\footnote{\url{gowers.wordpress.com/2022/04/28/announcing-an-automatic-theorem-proving-project/} (accessed June 2022)} to a project he initiated in 2008\footnote{\url{gowers.wordpress.com/2008/07/28/more-quasi-automatic-theorem-proving/}  (accessed June 2022)} to produce a ``human-style'' automated theorem-proving program \cite{GaneGowe17}. Here he means one that follows patterns of reasoning that appear in the proofs of human mathematicians, and with no exhaustive search (of say possible proofs through the space of proofs in some language). Gowers is especially interested in programs that would be useful to mathematicians, and is keen on output that actually \textit{reads} like proofs written by human mathematicians. That said, he seems more interested in program-generated proofs that provide \textit{insight} rather than just guarantee truth:
\begin{quote}
So what would be beneficial to ‘typical’ mathematicians? One possible answer is to place far less emphasis on proofs as certifications of correctness and far more on proofs as explanations. Many mathematicians say that their aim, when looking for proofs, is to achieve understanding rather than to feel confident that a statement is true.  \ldots

Therefore, for an automatic theorem prover to be useful to mainstream mathematicians, it will be highly desirable for it to produce ‘good’ proofs that ‘explain what is going on’ \cite[p.255]{GaneGowe17}.
\end{quote}

Furthermore ``human-style'' ATP must also include the design principle of working directly in the domains of the various mathematical sub-fields, with the objects (integers, matrices, graphs, etc) of those sub-fields, without translation to any more fundamental sub-field (for instance, set theory), just as human mathematicians do. 

By a ``domain'' here is meant a specific kind of object, all the properties defined for those objects, together with universal and existential statements quantified over those objects. The mathematical sub-field of \textit{graph theory} includes not just the domain of graphs, but basically anything graph-related including, for instance, claims about families of graphs (for instance, the Graph Minor Theorem). So while the Robertson-Seymour (AKA  Graph Minor) Theorem obviously is a theorem of graph theory broadly described, it is important to be clear that it is \textit{not} a statement quantified over graphs---it is quantified over \textit{families of graphs} \cite{Dies05}.

%

We will describe one possible top-down approach utilizing the \textsc{conjecturing} program which can produce conjectured necessary conditions for the objects in any domain to have a given property. \cite{LarsVanc17} For instance it can be used to generate necessary condition conjectures for a graph to have property $P$. We return to the proposed proof of Dirac's Theorem above: $\forall x [P(x)\rightarrow R(x)]$, $\forall x [(P(x) \& R(x)) \rightarrow S(x)]$, $\forall x [(P(x) \& R(x) \& S(x)) \rightarrow Q(x)]$, where we then conclude: $\forall x [P(x)\rightarrow Q(x)]$. The \textsc{conjecturing} program can generate necessary condition conjectures for graphs $x$ where $P(x)$ holds \cite{LarsVanc17}. It does this by generating expressions representing possible predicates. The simplest (atomic) ones are conditions that code graph theoretic properties. More complicated ones are built up from these as boolean functions of atomic predicates. Furthermore, there is a ``truth'' constraint on expression output: a proposed predicate $R(x)$ (and corresponding conjecture $\forall x [P(x)\rightarrow R(x)]$) can only be produced if the claim $P(x)\rightarrow R(x)$ is in fact true for all stored graphs $x$ (that is, if the collection of stored graphs $x$ that satisfy $P(x)$ are all contained in the collection of stored graphs $x$ that satisfy $R(x)$). It is worth noting that if Dirac's Theorem is true it must be the case that, for every stored graph $x$, if $P(x)$ is true then $Q(x)$ must also be true. 

Thus, a process that will yield this proof of Dirac's Theorem will first generate the necessary condition $R(x)$ for graphs $x$ where $P(x)$ holds. What might be wanted is that $R(x)$ is not only true for every stored graph $x$ where $P(x)$ is true but that there are a minimal number of stored graphs $x$ where $R(x)$ is true but $P(x)$ is not true (``minimality'' is intentionally undefined here---but can be addressed heuristically). Note, in this case, that it follows that for every stored graph $x$, if $P(x)\& R(x)$ is true then $Q(x)$ is true (this is just a consequence of the fact that, for every stored graph $x$, if $P(x)$ is true then $Q(x)$ must also be true). In this case we have a proof line: $\forall x [P(x)\rightarrow R(x)]$.

It might then be possible to iterate the above procedure. We now try to generate a necessary condition $S(x)$ for graphs $x$ where $P(x)\&R(x)$ holds. Again we want $S(x)$ to be  true for every stored graph $x$ where $P(x)\& R(x)$ is true but that there are a minimal number of stored graphs $x$ where $S(x)$ is true but $P(x)\&R(x)$ is not true. In this case we have another proof line: $\forall x [(P(x) \& R(x)) \rightarrow S(x)]$.

We will then continue to iterate the procedure as many times as possible (this is enforced to be finite in the \textsc{conjecturing} program by including a \textit{timeout}. Of course, from any number of atomic predicates, complex predicates of arbitrary size can be built-up from boolean operators---but conjectures must, in order to be comprehended and investigated, be of some human-readable, human-comprehendible, length.) At some point our program won't be able to produce a new predicate that meets our conditions within our time constraint, and outputs the last proof line: $\forall x [(P(x) \& R(x) \& S(x)) \rightarrow Q(x)]$. This proof is designed to justify the conclusion:  $\forall x [P(x)\rightarrow Q(x)]$. Whether it does, depends on the human that reads the proof. It may be incorrect (a proof line may actually be false). The leaps from line to line may be too great and require more justification. In any of these cases, the human mathematician necessarily holds information that can improve future proofs: she may know a counterexample to a proof line, and if she is stumped by an inference will have questions that will motivate new concepts which can also be fed back to the program.

It is possible to include theoretical knowledge (theorems) in this procedure. Perhaps it is known that $\forall x [P(x)\rightarrow Q_1(x)]$, $\forall x [P(x)\rightarrow Q_2(x)]$,\ldots, $\forall x [P(x)\rightarrow Q_k(x)]$, and that the predicates $Q_1(x)$, $Q_2(x)$,\ldots, $Q_k(x)$ have been coded. In this case it is enough to prove: 
\[
\forall x [(P(x)\&Q_1(x)\&Q_2(x)\& \ldots \& Q_k(x))\rightarrow Q(x)].
\]
 If we let $P'(x)$ be the predicate $P(x)\&Q_1(x)\&Q_2(x)\& \ldots \& Q_k(x)$ then what we need to prove is: $\forall x [P'(x)\rightarrow Q(x)]$ and then proceed exactly as in the case where we were proving: $\forall x [P(x)\rightarrow Q(x)]$. The significant difference will only we in the produced proof: it will include more facts and thus perhaps be more convincing to other mathematicians---and make the produced proof more likely to be accepted as a valid proof.






It is clearly important for the described program to ``know'' a large number of graph properties---the more properties that are coded the greater the chance that the described predicates $R(x)$ and $S(x)$ will be produced (the same phenomenon occurs with the \textsc{Conjecturing} program---the more properties are coded for an object-type the more likely it is that the program will produce a conjecture). 
It is also important for the program to know examples of significant graphs, ones that help guide theory development. Historically the famous Petersen graph was an important example: Petersen introduced it as a counterexample to Tait's claim that a bridgeless 3-regular graph is 3-edge-colorable \cite{BiggLloyWils86}. Thus any further claims about the edge-colorability of cubic graphs must be responsive to the Petersen graph (that is, must hold for this example).
A top-down program should thus know a large number of properties as well as a large number of significant examples. Any good human graph theorist will also know these. 






It was mentioned above that the semantic content of conjectures in the \textsc{Conjecturing} program are richer than the proposition objects in proof formalization programs like \textsc{Lean}. It was mentioned that these conjecture-objects include evaluation for specific objects among their methods. It might also be mentioned that the invariant and property terms in these conjectures are themselves semantically rich. 
The concepts/properties/invariants can be viewed as name-function pairs: each concept comes with a corresponding function that computes a number (for invariants) or a boolean (for properties). This approach may be more human-like: to know what a mathematical statement \textit{means} is more than to know what the definitions of the terms in the statement are---but rather how the definition applies (or not) in specific cases. We wouldn't say that a student understands the concept of graph \textit{hamiltonicity} if she can only give the definition but cannot explain why some specific graph (maybe a complete graph with five vertices) is hamiltonian.

Given some property of interest, say the property of being hamiltonian for a graph, the \textsc{Conjecturing} program can then be asked to produce necessary conditions for being hamiltonian. A produced necessary condition $G$ is true at least for the graphs the program ``knows'' (the input graphs). We use Fajtlowicz's \textit{Dalmatian} heuristic for deciding what to store temporarily and to eventually produce (an important feature of this heuristic is that the number of possible stored or conjectured necessary conditions is no more than the number of input objects---as with humans not much is kept in memory---this is true too for the expressions: while many are generated very very few are stored).

In doing research with the \textsc{Conjecturing} program, the user is at every point required to either prove the conjecture or to find a counter-example (while either of these processes might be automated, and some experiments have been done, it is not a feature of the program). If say a necessary condition conjecture is proved it can be added to the program as ``knowledge'', which in turn forces the program to make ``better'' conjectures moving forward (again, ``better'' is meant in a precise sense).  And if the user finds a counterexample to a conjecture this can also be stored---and again future conjectures will necessarily be ``better''.

Given some concepts and some either mathematical or propositional operators the program systematically generates all expressions (not-atomic, or complex, properties/concepts) of complexity-1 (the atomic concepts themselves), complexity-2 (unary operators applied to atomic concepts), etc. 
This might seem non-human-centric or machine-learning oriented, but somehow, somewhere in the brain, expressions must be formed. It might be thought that massive search is ``not human''. So maybe less systematic (more intelligent) expression search might also be possible. A neat fact though is that the expression-generator in \textsc{Conjecturing} can generate all possible expressions (millions) up to any human-comprehendible length in no more than a few seconds on any modern laptop. 

\section{More Issues and Objections}

The idea that an ATP program might be designed using fast expression-generation might raise some additional issues.

\begin{enumerate}

\item \textbf{It might be claimed that humans don't build expressions or do massive search the way that the Conjecturing program does}.
There are at least two related issues here. Researchers might want to code a program to model human mathematical abilities \textit{for the reason} that this might be the best way to code mathematical abilities. If human brains don't in fact do massive search in doing the things that mathematicians do then there should be non-search ways to perform these abilities, presumably more cleverly, more efficiently, etc, than by searching. In fact, we don't know how human brains do the things that mathematicians do. How do they find the property $P'$ needed in a particular mathematical circumstance? In fact, some people, Turing included, have thought that search is essential to intelligence. In the same 1948 paper where Turing proposed mathematics as an initial field of study for machine intelligence, he also speculated that ``intellectual activity consists mainly in various kinds of search.'' \cite{Turi48} And it is also worth noting that we may have brain abilities beyond what we consciously realize: Kim Peek was a savant who had memorized the contents of more than 9,000 books (with access to their contents by page number) \cite{Tref05}; Peek's case suggests the possibility of brain structure, capacity and abilities that we haven't begun yet to understand.

A second issue is that researchers might want to code programs that do the things intelligent mathematicians do in order to \textit{better understand} how human mathematicians do them. Again this is neither here nor there. If we don't actually know how a human mathematician could produce a concept $P'$ in order to need the needs of some problem, excluding methods like search because it doesn't \textit{seem} like its what a human brain can possibly do, can only be limiting. Sure any non-search heuristics are worth exploring, but maybe it is the case that humans do something analogous to systematic search.

In fact, the \textsc{conjecturing} program does not face the problem of interminable search (which may be the underlying objection to search): it forms expressions from atomic properties up to complex properties of complexity that are still human-comprehendible very fast (in less than a second)

\item \textbf{It might be asked which concepts should be used?} 
There are reasons to make a variety of choices here, in order to satisfy a variety of goals. So there can't be a ``correct'' answer here. What would be interesting is more information, from the work of developers, of what the outcomes of a variety of design choices are: what works and what doesn't? Rather than make any a priori requirements, it might be better to use empirical evidence to inform design choices. 

In fact, any published proof is knowledge intensive---a graph theorist (or any research mathematician) will know lots of concepts. It would certainly be non-human-oriented to require a program to reinvent her own concepts. And there are infinitely many graph theoretic concepts (invariants, properties) that could be invented or created---but the ones that a mathematician will typically see in a proof aren't created from scratch---they are exactly ones that are already known or are built-up from these. 

 For the purposes of our \textsc{conjecturing} program,  the more concepts the program knows the better the conjectures are (here ``better'' is used in a precise sense: for generated upper bound conjectures, say, the greater the number of input objects where the minimum of the conjectured upper bounds equals the value of the target invariant; this is not only an empirical observation but also a logical necessity due to the design of the program).  The authors have experimented with coding specific graphs and graph theory concepts---to leverage for the production of better graph theory conjectures. They are available at:\\

\url{https://github.com/math1um/objects-invariants-properties}\\

\item \textbf{Many proofs require a new idea or a new concept. Where will these come from? It might be thought that automatic concept generation also needs to be included in the design of any automated theorem-proving program.} 
It is true that new concepts are invented and appear in the mathematical literature, sometimes with rather limited motivation, but often in response to attempts to solve or address specific difficulties. It doesn't seem to be the case that these are \textit{all} really logical functions of pre-set atomic properties. And where did these atomic properties in any specific mathematical domain come from? Are they necessarily all possible atomic properties? There is no obvious reason why this should be the case.

Some, and maybe most, published proofs use existing concepts. So a first-step theorem-proving program might just use existing concepts. If a proof in fact needs a new idea this program won't find a proof. So it won't be the best possible theorem-proving program, but if it can produce some proofs some of the time it would be an advance. A program with even more abilities might do more.

There has for instance been research on the automated generation of new mathematical concepts \cite{Lena79,Colt99}. This research has often been motivated by a goal of producing ``interesting'' concepts; it would be more useful in this context to somehow produce concepts that help advance existing mathematical goals: find a new concept $P'$ for instance...


\end{enumerate}

\section{Advancing Top-down ATP}

There are good reasons to believe that top-down ATP approaches might lead to programs that can occasionally prove an open conjecture in well-studied domains in mathematics. It may be a key fact that research on the development of top-down programs must be more faithful to the actual practice of mathematicians. And, at least, new approaches might invigorate bottom-up ATP research. Wos for instance claimed, ``Heavy and continued experimentation is crucial to solving many of the research problems that currently confront automated reasoning'' \cite{Wos88}.

What should be done and what issues have \textit{not} been addressed?
 It is an important fact that mathematics is a large communal enterprise. It is reasonable to assume that a successful ATP program will function as a single (super) mathematical agent within a community of mathematicians---it is these mathematicians that will function as the ultimate arbiters as to what counts as ``rigorous'' and what concepts and properties will be included in the programs reasoning, and what examples its theorizing must be responsive to. 

\begin{enumerate}
\item \textbf{Code lots of knowledge}. 
It would be useful to code every graph that's appeared in a graph theory paper, every invariant or property that's appeared in a graph theory paper, and and every theorem about graphs. (It is also important to be clear about which concepts apply to graphs as opposed to families of graphs, pairs of graphs, collections of vertices in a graph of specific kinds, etc---these are all related, but of importantly different types.) Coding this knowledge is obviously beyond the abilities of any one human, but as a disciplinary project could have many advantages. While there are some 10 million connected graphs of order no more than ten, maybe there are only a few thousand graphs that have appeared in graph theory papers. Coding and using only graphs that humans have produced is possible and might seem reasonable: these are the ones that humans have thought enough about to be significant in some way. Searching through a few thousand graphs, slowly growing over time in response to human interests, should remain feasible---and might arguably not count as massive search.

Now it might seem ad hoc to code hundreds or thousands of domain-specific concepts in order to generate potential proofs in a domain, but in fact human mathematicians don't start from scratch; they start with a lot of graph theoretic knowledge. If you want a machine to produce human-like proofs the machine probably should have human-like knowledge to start. This might seem reminiscent of Lenat's \textsc{Cyc} \cite{Lena95} or maybe the expert-systems paradigm, both of which tried to code a lot of human knowledge---but in those cases it wasn't clear how that coded knowledge would translate to desired outputs. 
 We propose that knowledge will be encoded as concepts that will appear and be used in a specific way, which by design will advance the output goals of our programs.


\item \textbf{Pay careful attention to actual mathematical practice.}. 
It is important to look at what mathematicians actually \textit{do} and produce, rather than try to fit all of their productions into a universal framework. Mathematicians prove things about graphs, integers, matrices, continuous functions, etc. We prove statements about some specific kind of mathematical object, and never statements say quantified over ``mathematical objects'' (viewed as an abstraction). There are many commonalities, things that we do in proving mathematical statements in any domain---but maybe we do some things differently in some domains. There's no reason to start with any requirements of how acceptable proofs \textit{should} be, but rather to see what \textit{are} counted as acceptable proofs, and then produce similar proofs. 

%

\item \textbf{Examples \textit{are} knowledge.}
Every mathematician knows this. Our theories must first of all fit these fundamental examples. It might seem that examples are irrelevant for automated theorem proving programs---they don't appear in proofs. Nevertheless our theorem statements must hold for the objects of investigation. These statements should be semantically active---they should be things that can be checked against examples---and thus it is useful to have some selection of examples. 
It might seem that, if a proof is thought of largely syntactically as sequences of statements that are validly inferred for initial statements, examples are irrelevant. But if inferential leaps are allowed, it is crucial to spot-check inferred statements: they must for instance be true for objects in their domains of quantification. A program can do these checks if there is a store of examples. And the best examples are ones that have proven theoretically relevant in previous research.

\item \textbf{Mathematics grows in response to \textit{interaction}.}
The most important interactions are when one mathematician makes an inferential leap that is not understood or is challenged by other mathematicians. Then more explanation is required---and possibly new concepts will be created and appealed to. Automated theorem proving programs should grow along with the human mathematics they are engaged with---among other things their conceptual base must grow along with the mathematical domains they are engaged in.

Most interesting theorems began life as conjectures that weren't immediately resolved. They were investigated and in many (or most) cases new concepts were formed. To prove something in a domain never means to prove it as it existed at a certain date (with only the concepts that had been published up to that date). It means to prove it with all available concepts, as the subject grows. A successful theorem-proving program should also be continually enriched with the concepts in a domain. 


%
%

\end{enumerate}



Mathematics is often a limiting case. In philosophy for instance, a theory of knowledge should explain how we have all kinds of knowledge---including mathematical knowledge. But mathematical knowledge is importantly different from other kinds of knowledge. But without accounting for mathematical knowledge, an important kind of knowledge,  a theory of knowledge is incomplete. Currently artificial intelligence has made interesting and important advances. Some people think that machines will soon be smarter than humans. Well, this might be generally true. But again, mathematics seems like an important feature of human intelligence---and one where artificial intelligence programs don't seem to be advancing much on humans. Again, Turing's vision of programs that can prove interesting and new mathematical conjectures is still unrealized. 

But why? The reason could be that researchers in these last 60-odd years have focused on a specific paradigm of automated theorem-proving programs---build up from simple facts using valid inference rules. This may be too distant from actual mathematical practice. 
Developing theorem-proving programs that make inferential leaps may be the way to make future progress. Here we have explained how a program might make inferential leaps that are in some sense reasonable and even justified. In practice valid inferences are far more varied than ones prescribed by logicians:
ultimately what counts as a valid inference is just what mathematicians take to be a valid inference.  ATP program developers should model these inferences. 

\bibliographystyle{plain}
\bibliography{../../larson.bib}

\def\cprime{$'$} \def\cprime{$'$} \def\cprime{$'$} \def\cprime{$'$}
\begin{thebibliography}{10}

\bibitem{BiggLloyWils86}
N.~Biggs, E.~K. Lloyd, and R.~J. Wilson.
\newblock {\em Graph Theory, 1736-1936}.
\newblock Clarendon Press, 1986.

\bibitem{Carr95}
L.~Carroll.
\newblock What the tortoise said to {A}chilles.
\newblock {\em Mind}, 4(14):278--280, 1895.

\bibitem{Colt99}
S.~Colton.
\newblock Refactorable numbers---a machine invention.
\newblock {\em Journal of Integer Sequences}, 2(99.1):2, 1999.

\bibitem{Dies05}
R.~Diestel.
\newblock {\em Graph theory}, volume 173 of {\em Graduate Texts in
  Mathematics}.
\newblock Springer-Verlag, Berlin, third edition, 2005.

\bibitem{Fajt87}
S.~Fajtlowicz.
\newblock On conjectures of {G}raffiti. {II}.
\newblock {\em Congr. Numer.}, 60:189--197, 1987.
\newblock Eighteenth Southeastern International Conference on Combinatorics,
  Graph Theory, and Computing (Boca Raton, Fla., 1987).

\bibitem{Fajt88a}
S.~Fajtlowicz.
\newblock On conjectures of {G}raffiti.
\newblock In {\em Proceedings of the {F}irst {J}apan {C}onference on {G}raph
  {T}heory and {A}pplications ({H}akone, 1986)}, volume~72, pages 113--118,
  1988.

\bibitem{Fajt88}
S.~Fajtlowicz.
\newblock On conjectures of {G}raffiti. {III}.
\newblock {\em Congr. Numer.}, 66:23--32, 1988.
\newblock Nineteenth Southeastern Conference on Combinatorics, Graph Theory,
  and Computing (Baton Rouge, LA, 1988).

\bibitem{Fajt90}
S.~Fajtlowicz.
\newblock On conjectures of {G}raffiti. {IV}.
\newblock In {\em Proceedings of the {T}wentieth {S}outheastern {C}onference on
  {C}ombinatorics, {G}raph {T}heory, and {C}omputing ({B}oca {R}aton, {FL},
  1989)}, volume~70, pages 231--240, 1990.

\bibitem{Fajt95}
S.~Fajtlowicz.
\newblock On conjectures of {G}raffiti. {V}.
\newblock In {\em Graph {T}heory, {C}ombinatorics, and {A}lgorithms, {V}ol.\ 1,
  2 ({K}alamazoo, {MI}, 1992)}, Wiley-Intersci. Publ., pages 367--376. Wiley,
  New York, 1995.

\bibitem{GaneGowe17}
M.~Ganesalingam and W.~T. Gowers.
\newblock A fully automatic theorem prover with human-style output.
\newblock {\em Journal of Automated Reasoning}, 58(2):253--291, 2017.

\bibitem{GelbOlms03}
B.~R. Gelbaum and J.~Olmsted.
\newblock {\em Counterexamples in analysis}.
\newblock Courier Corporation, 2003.

\bibitem{Gele59}
H.~L. Gelernter.
\newblock Realization of a geometry theorem proving machine.
\newblock In {\em IFIP {C}ongress}, pages 273--281, 1959.

\bibitem{Hart15}
K.~Hartnett.
\newblock Will computers redefine the roots of math?
\newblock {\em Quanta Magazine}, 19.

\bibitem{Laka76}
I.~Lakatos.
\newblock {\em Proofs and refutations: The logic of mathematical discovery}.
\newblock Cambridge university press, 1976.

\bibitem{LarsVanc16}
C.~E. Larson and N.~Van~Cleemput.
\newblock Automated conjecturing {I}: Fajtlowicz's {D}almatian heuristic
  revisited.
\newblock {\em Artificial Intelligence}, 231:17--38, 2016.

\bibitem{LarsVanc17}
C.~E. Larson and N.~Van~Cleemput.
\newblock Automated conjecturing {III}: Property-relations conjectures.
\newblock {\em Annals of Mathematics and Artificial Intelligence},
  81(3):315--327, 2017.

\bibitem{Lena79}
D.~B. Lenat.
\newblock On automated scientific theory formation: a case study using the {AM}
  program.
\newblock {\em Machine intelligence}, 9:251--286, 1979.

\bibitem{Lena95}
D.B. Lenat.
\newblock Cyc: A large-scale investment in knowledge infrastructure.
\newblock {\em Communications of the ACM}, 38(11):33--38, 1995.

\bibitem{Mccu97}
W.~McCune.
\newblock Solution of the {R}obbins problem.
\newblock {\em Journal of Automated Reasoning}, 19(3):263--276, 1997.

\bibitem{NeweSimo56}
A.~Newell and H.~A. Simon.
\newblock The {L}ogic {T}heory machine.
\newblock {\em IRE Transactions of Information Theory}, 2:61--79, 1956.

\bibitem{Orne20}
S.~Ornes.
\newblock How close are computers to automating mathematical reasoning.
\newblock {\em Quanta Magazine}, 2020.

\bibitem{RabeSzeg21}
M.~N. Rabe and C.~Szegedy.
\newblock Towards the automatic mathematician.
\newblock In {\em International Conference on Automated Deduction (CADE)},
  pages 25--37. Springer, 2021.

\bibitem{Reek95}
M.~Reek.
\newblock A top-down approach to teaching programming.
\newblock In {\em Proceedings of the twenty-sixth SIGCSE technical symposium on
  Computer science education}, pages 6--9, 1995.

\bibitem{Robi65}
J.~A. Robinson.
\newblock A machine-oriented logic based on the resolution principle.
\newblock {\em Journal of the ACM (JACM)}, 12(1):23--41, 1965.

\bibitem{NeweSimo58}
H.~A. Simon and A.~Newell.
\newblock Heuristic problem solving: The next advance in operations research.
\newblock {\em Operations Research}, 6(1):1--10, 1958.

\bibitem{Tref05}
D.~A. Treffert and D.~D. Christensen.
\newblock Inside the mind of a savant.
\newblock {\em Scientific American}, 293(6):108--113, 2005.

\bibitem{Turi48}
A.~Turing.
\newblock pages 395--432. Oxford University Press, (1948)-2004.

\bibitem{Voev14}
V.~Voevodsky.
\newblock The origins and motivations of univalent foundations.
\newblock {\em IAS: The Institute Letter}, Summer 2014:8--9, 2014.

\bibitem{Wang60}
H.~Wang.
\newblock Toward mechanical mathematics.
\newblock {\em IBM J. Res. Develop.}, 4:2--22, 1960.

\bibitem{RussWhit97}
A.~N. Whitehead and B.~Russell.
\newblock {\em Principia mathematica to* 56}, volume~2.
\newblock Cambridge University Press, 1997.

\bibitem{WilfZeil92}
H.~S. Wilf and D.~Zeilberger.
\newblock An algorithmic proof theory for hypergeometric (ordinary and “q”)
  multisum/integral identities.
\newblock {\em Inventiones mathematicae}, 108(1):575--633, 1992.

\bibitem{Wos88}
L.~Wos.
\newblock {\em Automated reasoning: 33 basic research problems}.
\newblock Prentice Hall, 1988.

\bibitem{Wu12}
W.-t. Wu.
\newblock {\em Mechanical theorem proving in geometries: Basic principles}.
\newblock Springer Science \& Business Media, 2012.

\end{thebibliography}
%

\end{document}